\documentclass[letterpaper,aps,prl,superscriptaddress,showpacs,floatfix,twocolumn]{revtex4}

\usepackage{graphicx}	

\usepackage{xspace}	

\newcommand{\pt}{\mbox{$p_T$}\xspace}

\newcommand{\rdau}{\mbox{$R_{d\rm{Au}}$}\xspace}
\newcommand{\rcp}{\mbox{$R_{\rm CP}$}\xspace}
\newcommand{\pp}{\mbox{$p$+$p$}\xspace}

\newcommand{\dau}{\mbox{{\it d}+{\rm Au}}\xspace}  
\newcommand{\ncol}[1]{\mbox{$\left< N_{\rm coll}(#1)\right>$}\xspace}

\newcommand{\midn}{\mbox{$\left|\eta\right|<0.35$}\xspace}

\newcommand{\muonn}{\mbox{$1.2<|\eta|<2.4$}\xspace}

\newcommand \sqsn{\mbox{$\sqrt{s_{_{NN}}}$}\xspace}
\newcommand{\jpsi}{\mbox{$J/\psi$}\xspace}  

\newcommand{\ccbar}{\mbox{$c\overline{c}$}\xspace}
\newcommand{\sbr}{\mbox{$\sigma_{br}$}\xspace}
\newcommand{\lrt}{\mbox{$\Lambda(r_T)$}\xspace}
\newcommand{\rt}{\mbox{$r_T$}\xspace}

\begin{document}



\title{Cold Nuclear Matter Effects on $J/\psi$ Yields as a Function of
Rapidity and Nuclear Geometry in $d+$A Collisions 
at $\sqrt{s_{_{NN}}}$=200 GeV }

\newcommand{\abilene}{Abilene Christian University, Abilene, Texas 79699, USA}
\newcommand{\acadsin}{Institute of Physics, Academia Sinica, Taipei 11529, Taiwan}
\newcommand{\banaras}{Department of Physics, Banaras Hindu University, Varanasi 221005, India}
\newcommand{\barc}{Bhabha Atomic Research Centre, Bombay 400 085, India}
\newcommand{\bnlcoll}{Collider-Accelerator Department, Brookhaven National Laboratory, Upton, New York 11973-5000, USA}
\newcommand{\bnlphys}{Physics Department, Brookhaven National Laboratory, Upton, New York 11973-5000, USA}
\newcommand{\caucr}{University of California - Riverside, Riverside, California 92521, USA}
\newcommand{\charlesczech}{Charles University, Ovocn\'{y} trh 5, Praha 1, 116 36, Prague, Czech Republic}
\newcommand{\chonbuk}{Chonbuk National University, Jeonju, 561-756, Korea}
\newcommand{\ciae}{China Institute of Atomic Energy (CIAE), Beijing, People's Republic of China}
\newcommand{\cns}{Center for Nuclear Study, Graduate School of Science, University of Tokyo, 7-3-1 Hongo, Bunkyo, Tokyo 113-0033, Japan}
\newcommand{\colorado}{University of Colorado, Boulder, Colorado 80309, USA}
\newcommand{\columbia}{Columbia University, New York, New York 10027 and Nevis Laboratories, Irvington, New York 10533, USA}
\newcommand{\czechtech}{Czech Technical University, Zikova 4, 166 36 Prague 6, Czech Republic}
\newcommand{\dapnia}{Dapnia, CEA Saclay, F-91191, Gif-sur-Yvette, France}
\newcommand{\debrecen}{Debrecen University, H-4010 Debrecen, Egyetem t{\'e}r 1, Hungary}
\newcommand{\elte}{ELTE, E{\"o}tv{\"o}s Lor{\'a}nd University, H - 1117 Budapest, P{\'a}zm{\'a}ny P. s. 1/A, Hungary}
\newcommand{\ewha}{Ewha Womans University, Seoul 120-750, Korea}
\newcommand{\fit}{Florida Institute of Technology, Melbourne, Florida 32901, USA}
\newcommand{\fsu}{Florida State University, Tallahassee, Florida 32306, USA}
\newcommand{\gsu}{Georgia State University, Atlanta, Georgia 30303, USA}
\newcommand{\hiroshima}{Hiroshima University, Kagamiyama, Higashi-Hiroshima 739-8526, Japan}
\newcommand{\ihepprot}{IHEP Protvino, State Research Center of Russian Federation, Institute for High Energy Physics, Protvino, 142281, Russia}
\newcommand{\illuiuc}{University of Illinois at Urbana-Champaign, Urbana, Illinois 61801, USA}
\newcommand{\instpasczech}{Institute of Physics, Academy of Sciences of the Czech Republic, Na Slovance 2, 182 21 Prague 8, Czech Republic}
\newcommand{\isu}{Iowa State University, Ames, Iowa 50011, USA}
\newcommand{\jinrdubna}{Joint Institute for Nuclear Research, 141980 Dubna, Moscow Region, Russia}
\newcommand{\jyvaskyla}{Helsinki Institute of Physics and University of Jyv{\"a}skyl{\"a}, P.O.Box 35, FI-40014 Jyv{\"a}skyl{\"a}, Finland}
\newcommand{\kek}{KEK, High Energy Accelerator Research Organization, Tsukuba, Ibaraki 305-0801, Japan}
\newcommand{\kfki}{KFKI Research Institute for Particle and Nuclear Physics of the Hungarian Academy of Sciences (MTA KFKI RMKI), H-1525 Budapest 114, POBox 49, Budapest, Hungary}
\newcommand{\korea}{Korea University, Seoul, 136-701, Korea}
\newcommand{\kurchatov}{Russian Research Center ``Kurchatov Institute", Moscow, Russia}
\newcommand{\kyoto}{Kyoto University, Kyoto 606-8502, Japan}
\newcommand{\labllr}{Laboratoire Leprince-Ringuet, Ecole Polytechnique, CNRS-IN2P3, Route de Saclay, F-91128, Palaiseau, France}
\newcommand{\lawllnl}{Lawrence Livermore National Laboratory, Livermore, California 94550, USA}
\newcommand{\losalamos}{Los Alamos National Laboratory, Los Alamos, New Mexico 87545, USA}
\newcommand{\lpc}{LPC, Universit{\'e} Blaise Pascal, CNRS-IN2P3, Clermont-Fd, 63177 Aubiere Cedex, France}
\newcommand{\lund}{Department of Physics, Lund University, Box 118, SE-221 00 Lund, Sweden}
\newcommand{\maryland}{University of Maryland, College Park, Maryland 20742, USA}
\newcommand{\mass}{Department of Physics, University of Massachusetts, Amherst, Massachusetts 01003-9337, USA }
\newcommand{\muenster}{Institut fur Kernphysik, University of Muenster, D-48149 Muenster, Germany}
\newcommand{\muhlenberg}{Muhlenberg College, Allentown, Pennsylvania 18104-5586, USA}
\newcommand{\myongji}{Myongji University, Yongin, Kyonggido 449-728, Korea}
\newcommand{\nagasaki}{Nagasaki Institute of Applied Science, Nagasaki-shi, Nagasaki 851-0193, Japan}
\newcommand{\newmex}{University of New Mexico, Albuquerque, New Mexico 87131, USA }
\newcommand{\nmsu}{New Mexico State University, Las Cruces, New Mexico 88003, USA}
\newcommand{\ornl}{Oak Ridge National Laboratory, Oak Ridge, Tennessee 37831, USA}
\newcommand{\orsay}{IPN-Orsay, Universite Paris Sud, CNRS-IN2P3, BP1, F-91406, Orsay, France}
\newcommand{\peking}{Peking University, Beijing, People's Republic of China}
\newcommand{\pnpi}{PNPI, Petersburg Nuclear Physics Institute, Gatchina, Leningrad region, 188300, Russia}
\newcommand{\riken}{RIKEN Nishina Center for Accelerator-Based Science, Wako, Saitama 351-0198, Japan}
\newcommand{\rikjrbrc}{RIKEN BNL Research Center, Brookhaven National Laboratory, Upton, New York 11973-5000, USA}
\newcommand{\rikkyo}{Physics Department, Rikkyo University, 3-34-1 Nishi-Ikebukuro, Toshima, Tokyo 171-8501, Japan}
\newcommand{\saispbstu}{Saint Petersburg State Polytechnic University, St. Petersburg, Russia}
\newcommand{\saopaulo}{Universidade de S{\~a}o Paulo, Instituto de F\'{\i}sica, Caixa Postal 66318, S{\~a}o Paulo CEP05315-970, Brazil}
\newcommand{\seoulnat}{Seoul National University, Seoul, Korea}
\newcommand{\stonybrkc}{Chemistry Department, Stony Brook University, SUNY, Stony Brook, New York 11794-3400, USA}
\newcommand{\stonycrkp}{Department of Physics and Astronomy, Stony Brook University, SUNY, Stony Brook, New York 11794-3400, USA}
\newcommand{\subatech}{SUBATECH (Ecole des Mines de Nantes, CNRS-IN2P3, Universit{\'e} de Nantes) BP 20722 - 44307, Nantes, France}
\newcommand{\tenn}{University of Tennessee, Knoxville, Tennessee 37996, USA}
\newcommand{\titech}{Department of Physics, Tokyo Institute of Technology, Oh-okayama, Meguro, Tokyo 152-8551, Japan}
\newcommand{\tsukuba}{Institute of Physics, University of Tsukuba, Tsukuba, Ibaraki 305, Japan}
\newcommand{\vandy}{Vanderbilt University, Nashville, Tennessee 37235, USA}
\newcommand{\waseda}{Waseda University, Advanced Research Institute for Science and Engineering, 17 Kikui-cho, Shinjuku-ku, Tokyo 162-0044, Japan}
\newcommand{\weizmann}{Weizmann Institute, Rehovot 76100, Israel}
\newcommand{\yonsei}{Yonsei University, IPAP, Seoul 120-749, Korea}
\affiliation{\abilene}
\affiliation{\acadsin}
\affiliation{\banaras}
\affiliation{\barc}
\affiliation{\bnlcoll}
\affiliation{\bnlphys}
\affiliation{\caucr}
\affiliation{\charlesczech}
\affiliation{\chonbuk}
\affiliation{\ciae}
\affiliation{\cns}
\affiliation{\colorado}
\affiliation{\columbia}
\affiliation{\czechtech}
\affiliation{\dapnia}
\affiliation{\debrecen}
\affiliation{\elte}
\affiliation{\ewha}
\affiliation{\fit}
\affiliation{\fsu}
\affiliation{\gsu}
\affiliation{\hiroshima}
\affiliation{\ihepprot}
\affiliation{\illuiuc}
\affiliation{\instpasczech}
\affiliation{\isu}
\affiliation{\jinrdubna}
\affiliation{\jyvaskyla}
\affiliation{\kek}
\affiliation{\kfki}
\affiliation{\korea}
\affiliation{\kurchatov}
\affiliation{\kyoto}
\affiliation{\labllr}
\affiliation{\lawllnl}
\affiliation{\losalamos}
\affiliation{\lpc}
\affiliation{\lund}
\affiliation{\maryland}
\affiliation{\mass}
\affiliation{\muenster}
\affiliation{\muhlenberg}
\affiliation{\myongji}
\affiliation{\nagasaki}
\affiliation{\newmex}
\affiliation{\nmsu}
\affiliation{\ornl}
\affiliation{\orsay}
\affiliation{\peking}
\affiliation{\pnpi}
\affiliation{\riken}
\affiliation{\rikjrbrc}
\affiliation{\rikkyo}
\affiliation{\saispbstu}
\affiliation{\saopaulo}
\affiliation{\seoulnat}
\affiliation{\stonybrkc}
\affiliation{\stonycrkp}
\affiliation{\subatech}
\affiliation{\tenn}
\affiliation{\titech}
\affiliation{\tsukuba}
\affiliation{\vandy}
\affiliation{\waseda}
\affiliation{\weizmann}
\affiliation{\yonsei}
\author{A.~Adare} \affiliation{\colorado}
\author{S.~Afanasiev} \affiliation{\jinrdubna}
\author{C.~Aidala} \affiliation{\mass}
\author{N.N.~Ajitanand} \affiliation{\stonybrkc}
\author{Y.~Akiba} \affiliation{\riken} \affiliation{\rikjrbrc}
\author{H.~Al-Bataineh} \affiliation{\nmsu}
\author{J.~Alexander} \affiliation{\stonybrkc}
\author{A.~Angerami} \affiliation{\columbia}
\author{K.~Aoki} \affiliation{\kyoto} \affiliation{\riken}
\author{N.~Apadula} \affiliation{\stonycrkp}
% ...
\collaboration{PHENIX Collaboration} \noaffiliation

\date{\today}

\begin{abstract}
  We present measurements of $J/\psi$ yields in $d+$Au collisions
  at $\sqrt{s_{_{NN}}}$=200 GeV recorded by the PHENIX experiment and 
  compare with
  yields in $p+p$ collisions at the same energy per
  nucleon-nucleon collision. The measurements cover a large kinematic
  range in $J/\psi$ rapidity ($-2.2 < y < 2.4$) with high statistical
  precision and are compared with two theoretical models: one with
  nuclear shadowing combined with final state breakup and one with
  coherent gluon saturation effects. In order to remove model
  dependent systematic uncertainties we also compare the data to a
  simple geometric model. We find that calculations where the nuclear
  modification is linear or exponential in the density weighted
  longitudinal thickness are difficult to reconcile with the forward
  rapidity data.

\end{abstract}

	


\maketitle

\label{sec:cnm}

The measured yields of quarkonia states in $p+$A (or $d+$A)  
collisions provide information about the time scale and dynamics 
for the creation of a \ccbar pair and its evolution to a 
color-singlet quarkonium state.  The propagation time of the 
\ccbar pair through the nucleus is set by the incident energy of 
the projectile and target and by the relative longitudinal 
momentum of the \ccbar pair. Fixed target $p+$A experiments at 
Fermilab~\cite{Leitch:1999ea} reveal a substantial suppression for 
forward rapidity \jpsi and $\psi'$ at a similar level, leading to 
the conclusion that the suppression must occur at the prehadronic 
stage.  An analysis~\cite{Lourenco:2008sk} of results for 
\sqsn= 17--42 GeV indicates that in 
addition to modified initial production due to nuclear-modified 
parton distribution functions (nPDFs), a break up cross section 
(\sbr) for the \ccbar precursor state to the \jpsi is important, 
and that \sbr decreases as the relative center-of-mass energy 
between the \ccbar and the nucleon increases.  Extending these 
results to collider energies at RHIC is important. The dominant 
production mechanism for charm (at RHIC) is via gluon-gluon 
interactions, and thus the yields at forward rapidity, the 
deuteron-going direction, are sensitive to low-$x$ in the gold 
nucleus where gluon 
shadowing~\cite{deFlorian:2003qf,Eskola:2001gt} and/or gluon 
saturation effects~\cite{Kharzeev:2005zr} become important, 
providing a crucial test of these effects.

There is also significant interest in determining the color screening
length in the quark-gluon plasma for temperatures $T > 170$ MeV, as
achieved in relativistic heavy ion
collisions~\cite{Adare:2008fqa}. One proposal for determining this is
the measurement of several quarkonia states where the different
binding energies (and thus radii) can bracket the screening length of
interest~\cite{Matsui:1986dk,Karsch:2005nk}. However, this suppression
of quarkonia must be separated from the aforementioned cold nuclear
matter effects. Thus precise measurement of quarkonia suppression in
$d+$Au is needed.

The PHENIX experiment has previously published \jpsi results in
$d+$Au collisions at \sqsn = 200 GeV~\cite{Adare:2007gn} from
data taken in 2003. In this paper we present results from
$d+$Au collision data taken in 2008, representing an increase
in yield by a factor of 30--50 over the previous results and a
reduction in the systematic uncertainties by up to a factor of
two.  Additionally, the $p+p$ reference data sets are updated
to include higher statistics data from 2006 and 2008.

The PHENIX apparatus is described in detail in~\cite{Adcox:2003zm}. It
consists of two sets of spectrometers referred to as the central arms,
which measure single-particles emitted over pseudorapidity ($\midn$),
and the muon arms, measuring single muons over pseudorapidities
($\muonn$). \jpsi particles are measured via their dielectron
(dimuon) decays at mid (backward and forward) rapidities, and
detailed analysis methods are given in~\cite{Adare:2007gn,PPG104}. The
\dau data used for this analysis were recorded using selective Level-1
triggers in coincidence with a minimum bias interaction requirement,
which requires one hit in each of two beam-beam counters (BBCs)
located at positive and negative pseudorapidity
($3<|\eta|<3.9$). This minimum bias selection covers $88 \pm 4$\% of
the total \dau inelastic cross section of 2260
mb~\cite{White:2005kp}. This can be corrected to an unbiased sample,
$100$\% of the total cross section, by applying a bias correction
factor ($c$ = 0.89) to the particle yield measured in any minimum bias
event. 
Additional Level-1 triggers independently require (1) one hit above 
threshold (600 or 800 MeV) in the Electromagnetic Calorimeter with a 
matching hit in the Ring Imaging \v{C}erenkov Detector identified 
as an electron or (2) two tracks identified as muon 
candidates~\cite{Adare:2007gn}.
The data sets
sampled via the Level-1 triggers represent analyzed integrated
luminosities for the different spectrometers of 54.0 nb$^{-1}$ to 69.3
nb$^{-1}$.  We use $p+p$ reference data for the midrapidity
dielectrons from~\cite{Adare:2009js}.    For the forward and backward 
rapidity dimuons, we report here new $p+p$ data from 2006 and 2008 
with a total integrated luminosity of 5.1 pb$^{-1}$.

\begin{figure}[thb]
  \centering
  \includegraphics[width=0.9\linewidth]{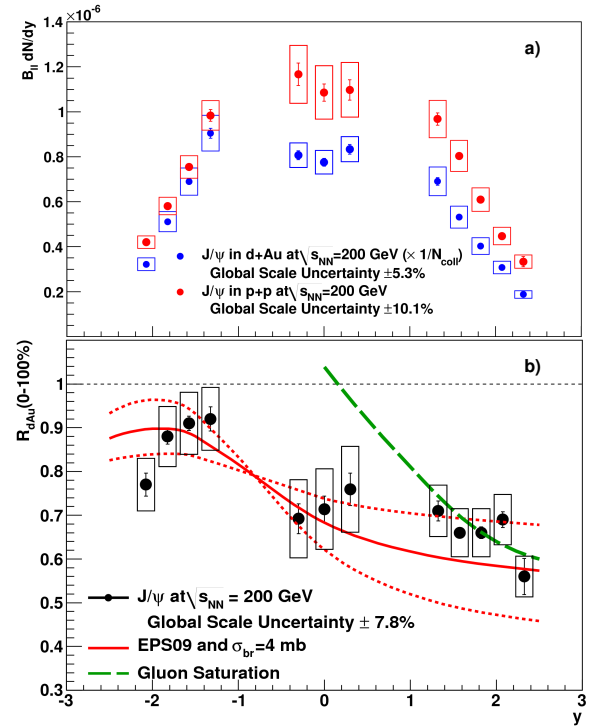} 
  \caption{\label{fig:invyrap} (color online) (a) $\jpsi$ $B_{ll}$ $dN/dy$
    in \pp and \dau collisions as a function of rapidity.  The \dau yields 
are divided by the average number of nucleon-nucleon collisions  
\ncol{0\mbox{-}100\%} = 7.6.
(error lines) point-to-point uncorrelated
uncertainties; (boxes) point-to-point
correlated uncertainties; (text) global
normalization scale uncertainties.
(b)  \rdau nuclear modification factors for unbiased collisions. 
}
\end{figure}

The \pt-integrated \jpsi yield as a function of rapidity is calculated via:
\begin{equation}
  \centering B_{ll} \frac{dN}{dy} = \frac{c N_{J/\psi}}{N_{\rm MB} \epsilon A \Delta{y}}
\label{eq:invy}
\end{equation}
where $B_{ll}$ is the branching fraction for \jpsi $\rightarrow
e^{+}e^{-}$ or $\mu^{+}\mu^{-}$, $N_{J/\psi}$ is the number of \jpsi
counts, $c$ is the bias correction factor, $N_{\rm MB}$ is the number of
sampled minimum bias events, $\Delta{y}$ is the width of the rapidity
bin, and $\epsilon A$ represents the product of the efficiency and
acceptance corrections, including the Level-1 trigger efficiency.  The
number of \jpsi particles is determined using the invariant mass
distribution of unlike-sign lepton pairs. Approximately 38000, 8900,
and 42000 \jpsi counts are measured at backward, mid, and forward
rapidity, respectively.  Figure~\ref{fig:invyrap}a shows the \jpsi yields in \pp and
\dau unbiased collisions.  

\begin{figure}[th]
  \includegraphics[width=0.9\linewidth]{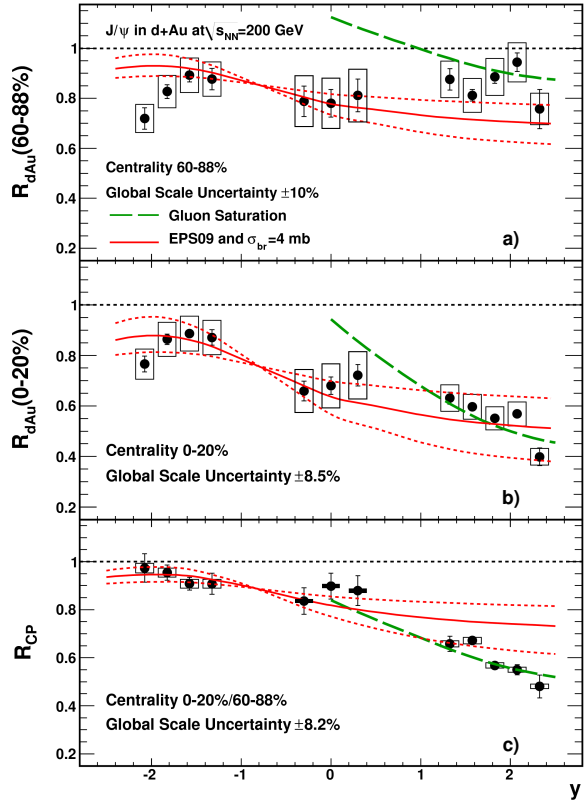}
  \caption{\label{fig:rcp} (color online) Nuclear suppression factors
    (\rdau) for (a) peripheral, (b) central, and (c) \rcp as a
    function of rapidity.  
}
\end{figure}

We quantify the cold nuclear matter effects by calculating the nuclear modification
factor \rdau as given by:
\begin{equation}
\centering \rdau(i) = \frac {\frac{dN^{d+{\rm Au}}(i)}{dy}} {\ncol{i}
\frac{dN^{p+p}}{dy}}
\label{eq:rdau}
\end{equation}
where $i$ is the index of the centrality bin and \ncol{i} is the
average number of nucleon-nucleon collisions and is determined from
the total energy deposited in the BBC located at negative rapidity.
For a given centrality bin \ncol{i} is derived using a Glauber
calculation coupled to a simulation of the BBC response (as described
in ~\cite{Adare:2007gn}). The centrality bins used in this analysis
are characterized as follows: central \ncol{0\mbox{--}20\%} = 15.1
$\pm$ 1.0, \ncol{20\mbox{--}40\%} = 10.3 $\pm$ 0.7,
\ncol{40\mbox{--}60\%} = 6.6 $\pm$ 0.6, \ncol{60\mbox{--}88\%} = 3.2
$\pm$ 0.2 and unbiased \ncol{0\mbox{--}100\%} = 7.6 $\pm$
0.3. Figure~\ref{fig:invyrap}b shows the \rdau corresponding to
unbiased collisions. Figure~\ref{fig:rcp}a (b) shows \rdau
corresponding to \dau centralities of 60--88\% (0--20\%). Note that more
central collisions correspond to cases where the nucleons in the
deuteron strike closer to the middle of the gold nucleus, and thus
nuclear effects are expected to be enhanced (which is seen in the
data).

The peripheral \rdau favors some suppression at all rapidities, though
this result is tempered by the current systematics of approximately
$\pm$15\%.  The central \rdau indicates a much larger suppression for
$J/\psi$ at forward rapidity.

We also calculate the ratio \rcp as the nuclear modification between
central and peripheral $d+$Au collision classes of events:
\begin{equation}
\label{eq:rcp}
\centering \rcp = \frac {\frac{dN^{d+{\rm Au}}(0\mbox{--}20\%)}{dy} /
\ncol{0\mbox{--}20\%}}  {\frac{dN^{d+{\rm Au}}(60\mbox{--}88\%)}{dy} /
\ncol{60\mbox{--}88\%}}
\end{equation}

Figure~\ref{fig:rcp}c shows the \rcp ratio for the most
central category relative to the peripheral 60--88\% category as a
function of rapidity. The quantity \rcp has the advantage that many of
the systematic uncertainties cancel in the ratio.  One observes a
dramatic suppression of forward rapidity yields for central \dau
events compared to peripheral events. At backward rapidity, there is
little to no modification seen.

We confront two specific calculations available in the literature with
our data. The first class of calculations that are often employed
include nuclear-modified PDFs and an effective \sbr. Here we utilize
the EPS09 nuclear modified PDF set~\cite{Eskola:2009uj} and a $\sbr=4$
mb (chosen to match the unbiased backward rapidity \rdau
data)~\cite{Vogt:2004dh}.  We find a reasonable agreement within
uncertainties with the unbiased \rdau data shown in
Fig.~\ref{fig:invyrap}b. We also show as red dashed lines the
differences within the EPS09 nPDFs for a single parameter change that
gives the largest variation~\cite{Eskola:2009uj}.  However, if one
plots the same calculations for \rcp as shown in
Fig.~\ref{fig:rcp}c, one sees reasonable agreement at backward and
midrapidities, but a significant under-prediction of the suppression
for forward rapidity \jpsi in the most central events.

A second class of calculations incorporates gluon saturation effects
at small-$x$~\cite{Kharzeev:2005zr,Kharzeev:2003sk}, and is compared
with experimental data in Figs.~\ref{fig:invyrap}-\ref{fig:rcp}.  A
modest \jpsi enhancement is predicted at midrapidity due to
double-gluon exchange processes (not seen in the data) and a
substantial \jpsi suppression at forward rapidity and in more central
\dau events due to saturation effects (in agreement with the data).

\begin{figure}[th]
  \includegraphics[width=0.9\linewidth]{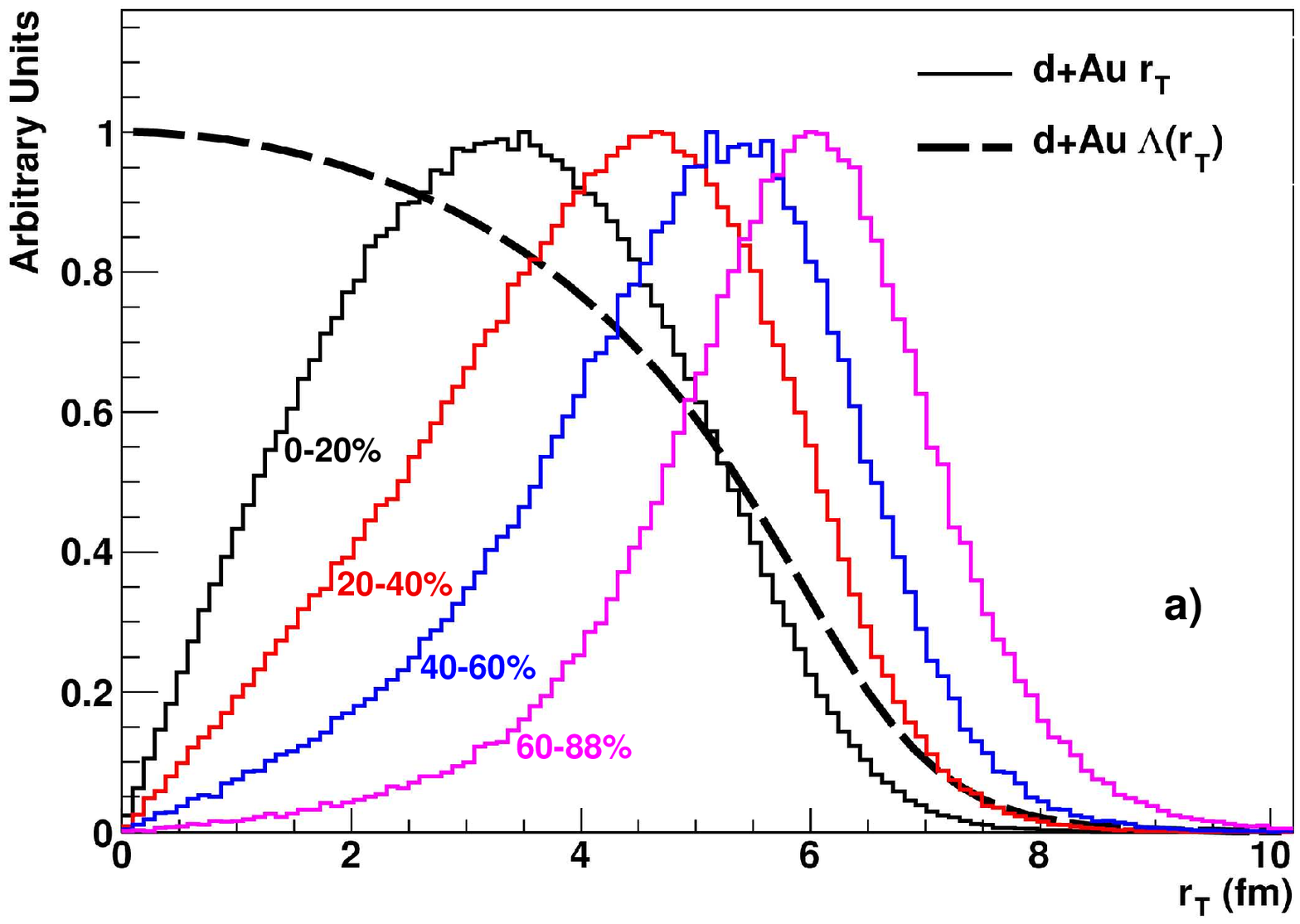} 
  \includegraphics[width=0.9\linewidth]{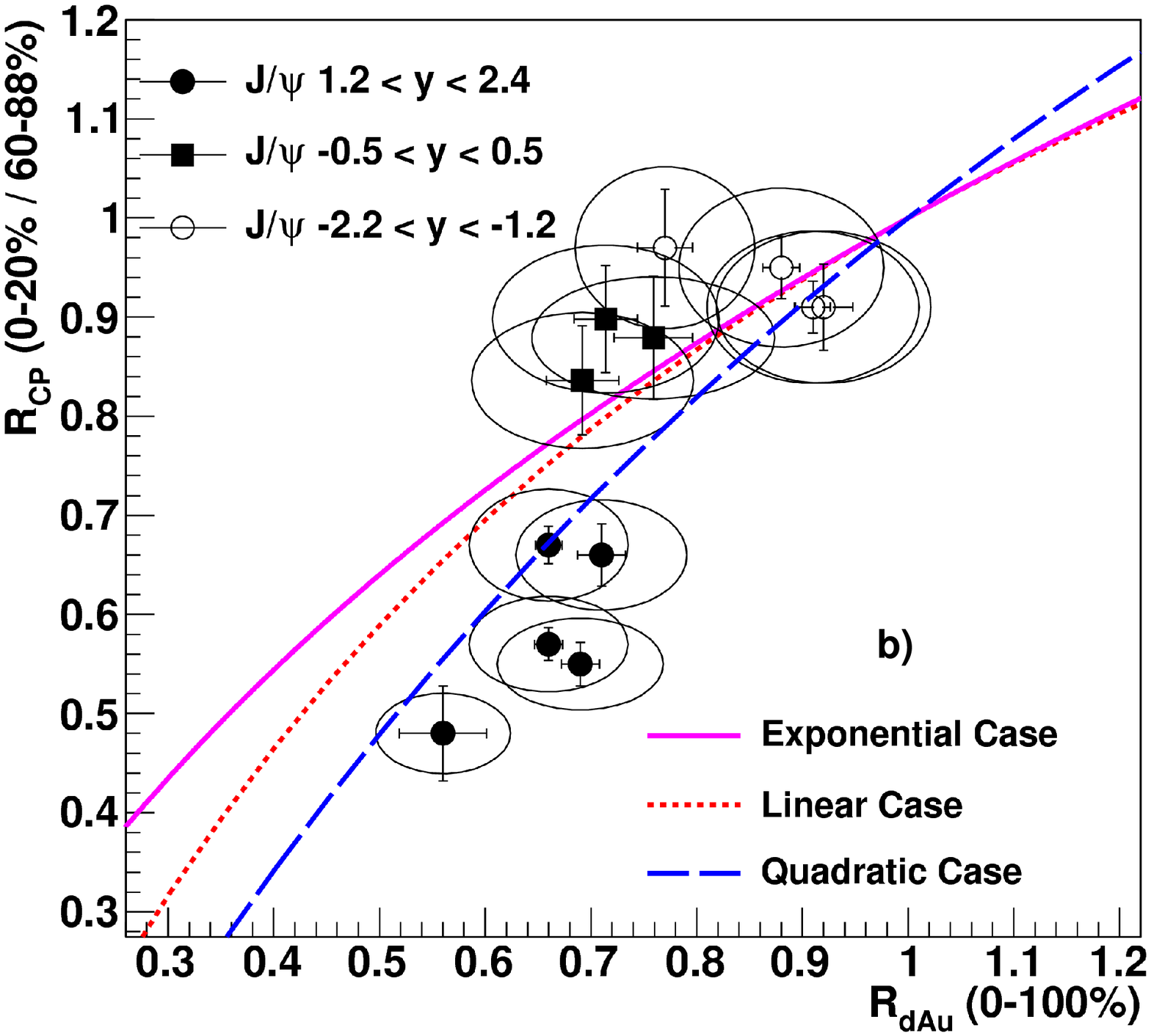} 
  \caption{\label{fig:linear} (color online) (a) 
Normalized to unity at the maximum bin are (solid curves) transverse radial
    \rt distributions in the gold nucleus for four \dau
    centrality selections and (dashed curve) density weighted
    longitudinal thickness as a function of \rt (\lrt).  
(b) (points) Unbiased \rdau versus \rcp for the experimental data and (curves)
    constraint lines for three geometric dependencies of the
    nuclear modification.}
\end{figure}

In order to further explore the centrality dependence of the nuclear
effects we categorize each \dau centrality class in terms of the
distribution of transverse radial positions (\rt) of the nucleon-nucleon
collisions relative to the center of the gold nucleus. The \rt
distributions for the four centrality categories are shown in
Fig.~\ref{fig:linear}a. We expect that the nuclear
effects are dependent on the density weighted longitudinal thickness
through the gold nucleus $\left(\lrt \equiv {\frac {1}{\rho_{0}}} \int dz
\rho(z,r_{T})\right)$, where $\rho_{0}$ is the density in the middle of the
nucleus. This quantity is also shown in Fig.~\ref{fig:linear}a as a
function of \rt.

We now posit three different functional dependencies of the nuclear
modification on \lrt.
\begin{eqnarray}
\label{eq:exp}
{\rm Exponential}:  M(r_{T}) = e^{ -a\lrt} \\
\label{eq:lin}
{\rm Linear}:  M(r_{T}) = 1.0 - a\lrt \\
\label{eq:quad}
{\rm Quadratic}: M(r_{T}) = 1.0 - a\lrt^2,
\end{eqnarray}
where $a$ is a parameter depending on the average level of
modification. The EPS09 nPDF based calculation, shown in
Figs.~\ref{fig:invyrap}~\ref{fig:rcp}, assumes the linear
relation~\cite{Klein:2003dj,Vogt:2004dh} in Eq.~\ref{eq:lin} in order
to make centrality-dependent predictions. In contrast, contributions
from a break up of the \ccbar via a \sbr follow the exponential
relation in Eq.~\ref{eq:exp}.

Figure~\ref{fig:linear}b shows the nuclear modification \rcp in
the most central bin versus the (unbiased) average modification \rdau.
This particular set of quantities is chosen because for each of the
three geometric dependencies
(Eqs.~\ref{eq:exp}--\ref{eq:quad}), a given value of the
parameter $a$ results in a unique point on the plot and varying the
parameter $a$ results in a unique locus of points on which any suppression
with that geometric dependence must lie.

The experimental data is also plotted in Fig.~\ref{fig:linear}b for
the same quantities.  The ellipses represent a one standard deviation
contour for the systematic uncertainties, which are
largely uncorrelated between the unbiased \rdau and \rcp.  There is a
substantial deviation between the exponential and linear cases and the
experimental data at forward rapidity, while at mid and backward
rapidities the data cannot discriminate between the cases. Thus, the
forward rapidity data suggest that the dependence on \lrt is
non-linear and closer to quadratic. If the dominant physics leading to
the modification is different at different rapidities, it is possible
for example that the modification at backward rapidities is linear
while at forward rapidities it is not. This is reinforced by the EPS09
plus \sbr calculation where regardless of the variation of the nPDF or
\sbr one cannot simultaneously describe the full centrality dependence
of the data as seen in Fig.~\ref{fig:rcp}.

Other non-linear density effects (\textit{e.g.}, quadratic) for the
geometric dependence~\cite{Frankfurt:2003zd} and for break up of the
\ccbar after production~\cite{Qiu:1998rz,Kopeliovich:2010nw} have been
proposed. An alternative explanation is that initial-state parton
energy loss results in a backward shift of the \jpsi rapidity
distribution~\cite{Johnson:2001xfa}.  It has been
observed~\cite{LindenLevy:2009td} that the nuclear modification as a
function of center-of-mass rapidity is similar to that observed at
lower energies~\cite{Leitch:1999ea} with a steep increase in
suppression at forward rapidities, as predicted for initial-state
parton energy loss.

In summary, we have presented precision data on \jpsi yields in \dau
and \pp collisions at \sqsn = 200 GeV over a broad range in rapidity
and \dau centrality. Nuclear modification factors at forward rapidity
as a function of centrality cannot be reconciled with a picture of
cold nuclear matter effects (nPDFs and a \sbr) when an exponential or
linear dependence on the nuclear thickness is employed. Effects of
gluon saturation may play an important role in understanding the
forward rapidity modifications, though other explanations involving
initial-state parton energy loss need further
investigation. 


\begin{acknowledgments}

We thank the staff of the Collider-Accelerator and
Physics Departments at BNL for their vital contributions.
We also thank Ramona Vogt and Kirill
Tuchin for useful discussions and theoretical calculations.
We acknowledge support from
the Office of Nuclear Physics in DOE Office of Science, NSF,
and a sponsored research grant from Renaissance Technologies (USA),
MEXT and JSPS (Japan),
CNPq and FAPESP (Brazil),
NSFC (China),
MSMT (Czech Republic),
IN2P3/CNRS and CEA (France),
BMBF, DAAD, and AvH (Germany),
OTKA (Hungary),
DAE and DST (India),
ISF (Israel),
NRF and WCU (Korea),
MES, RAS, and FAAE (Russia),
VR and KAW (Sweden),
U.S. CRDF for the FSU,
US-Hungary Fulbright,
and US-Israel BSF.

\end{acknowledgments}



\begin{thebibliography}{22}
\expandafter\ifx\csname natexlab\endcsname\relax\def\natexlab#1{#1}\fi
\expandafter\ifx\csname bibnamefont\endcsname\relax
  \def\bibnamefont#1{#1}\fi
\expandafter\ifx\csname bibfnamefont\endcsname\relax
  \def\bibfnamefont#1{#1}\fi
\expandafter\ifx\csname citenamefont\endcsname\relax
  \def\citenamefont#1{#1}\fi
\expandafter\ifx\csname url\endcsname\relax
  \def\url#1{\texttt{#1}}\fi
\expandafter\ifx\csname urlprefix\endcsname\relax\def\urlprefix{URL }\fi
\providecommand{\bibinfo}[2]{#2}
\providecommand{\eprint}[2][]{\url{#2}}

\bibitem[{\citenamefont{Leitch et~al.}(2000)}]{Leitch:1999ea}
\bibinfo{author}{\bibfnamefont{M.~J.} \bibnamefont{Leitch}}
  \bibnamefont{et~al.} (\bibinfo{collaboration}{FNAL E866/NuSea
  Collaboration}), \bibinfo{journal}{Phys. Rev. Lett.}
  \textbf{\bibinfo{volume}{84}}, \bibinfo{pages}{3256} (\bibinfo{year}{2000}).

\bibitem[{\citenamefont{Lourenco et~al.}(2009)\citenamefont{Lourenco, Vogt, and
  Woehri}}]{Lourenco:2008sk}
\bibinfo{author}{\bibfnamefont{C.}~\bibnamefont{Lourenco}},
  \bibinfo{author}{\bibfnamefont{R.}~\bibnamefont{Vogt}}, \bibnamefont{and}
  \bibinfo{author}{\bibfnamefont{H.~K.} \bibnamefont{Woehri}},
  \bibinfo{journal}{JHEP} \textbf{\bibinfo{volume}{02}}, \bibinfo{pages}{014}
  (\bibinfo{year}{2009}), \bibinfo{note}{and references therein.}

\bibitem[{\citenamefont{de~Florian and Sassot}(2004)}]{deFlorian:2003qf}
\bibinfo{author}{\bibfnamefont{D.}~\bibnamefont{de~Florian}} \bibnamefont{and}
  \bibinfo{author}{\bibfnamefont{R.}~\bibnamefont{Sassot}},
  \bibinfo{journal}{Phys. Rev.} \textbf{\bibinfo{volume}{D69}},
  \bibinfo{pages}{074028} (\bibinfo{year}{2004}).

\bibitem[{\citenamefont{Eskola et~al.}(2001)\citenamefont{Eskola, Kolhinen, and
  Vogt}}]{Eskola:2001gt}
\bibinfo{author}{\bibfnamefont{K.~J.} \bibnamefont{Eskola}},
  \bibinfo{author}{\bibfnamefont{V.~J.} \bibnamefont{Kolhinen}},
  \bibnamefont{and} \bibinfo{author}{\bibfnamefont{R.}~\bibnamefont{Vogt}},
  \bibinfo{journal}{Nucl. Phys.} \textbf{\bibinfo{volume}{A696}},
  \bibinfo{pages}{729} (\bibinfo{year}{2001}).

\bibitem[{\citenamefont{Kharzeev and Tuchin}(2006)}]{Kharzeev:2005zr}
\bibinfo{author}{\bibfnamefont{D.}~\bibnamefont{Kharzeev}} \bibnamefont{and}
  \bibinfo{author}{\bibfnamefont{K.}~\bibnamefont{Tuchin}},
  \bibinfo{journal}{Nucl. Phys.} \textbf{\bibinfo{volume}{A770}},
  \bibinfo{pages}{40} (\bibinfo{year}{2006}).

\bibitem[{\citenamefont{Adare et~al.}(2010{\natexlab{a}})}]{Adare:2008fqa}
\bibinfo{author}{\bibfnamefont{A.}~\bibnamefont{Adare}} \bibnamefont{et~al.}
  (\bibinfo{collaboration}{PHENIX Collaboration}), \bibinfo{journal}{Phys. Rev.
  Lett.} \textbf{\bibinfo{volume}{104}}, \bibinfo{pages}{132301}
  (\bibinfo{year}{2010}{\natexlab{a}}).

\bibitem[{\citenamefont{Matsui and Satz}(1986)}]{Matsui:1986dk}
\bibinfo{author}{\bibfnamefont{T.}~\bibnamefont{Matsui}} \bibnamefont{and}
  \bibinfo{author}{\bibfnamefont{H.}~\bibnamefont{Satz}},
  \bibinfo{journal}{Phys. Lett.} \textbf{\bibinfo{volume}{B178}},
  \bibinfo{pages}{416} (\bibinfo{year}{1986}).

\bibitem[{\citenamefont{Karsch et~al.}(2006)\citenamefont{Karsch, Kharzeev, and
  Satz}}]{Karsch:2005nk}
\bibinfo{author}{\bibfnamefont{F.}~\bibnamefont{Karsch}},
  \bibinfo{author}{\bibfnamefont{D.}~\bibnamefont{Kharzeev}}, \bibnamefont{and}
  \bibinfo{author}{\bibfnamefont{H.}~\bibnamefont{Satz}},
  \bibinfo{journal}{Phys. Lett.} \textbf{\bibinfo{volume}{B637}},
  \bibinfo{pages}{75} (\bibinfo{year}{2006}).

\bibitem[{\citenamefont{Adare et~al.}(2008)}]{Adare:2007gn}
\bibinfo{author}{\bibfnamefont{A.}~\bibnamefont{Adare}} \bibnamefont{et~al.}
  (\bibinfo{collaboration}{PHENIX Collaboration}), \bibinfo{journal}{Phys.
  Rev.} \textbf{\bibinfo{volume}{C77}}, \bibinfo{pages}{024912}
  (\bibinfo{year}{2008}).

\bibitem[{\citenamefont{Adcox et~al.}(2003)}]{Adcox:2003zm}
\bibinfo{author}{\bibfnamefont{K.}~\bibnamefont{Adcox}} \bibnamefont{et~al.}
  (\bibinfo{collaboration}{PHENIX Collaboration}), \bibinfo{journal}{Nucl.
  Instrum. Meth.} \textbf{\bibinfo{volume}{A499}}, \bibinfo{pages}{469}
  (\bibinfo{year}{2003}).

\bibitem[{PPG()}]{PPG104}
\bibinfo{note}{A. Adare et al. (PHENIX Collaboration) to be published.}

\bibitem[{\citenamefont{White}(2005)}]{White:2005kp}
\bibinfo{author}{\bibfnamefont{S.~N.} \bibnamefont{White}},
  \bibinfo{journal}{AIP Conf. Proc.} \textbf{\bibinfo{volume}{792}},
  \bibinfo{pages}{527} (\bibinfo{year}{2005}).

\bibitem[{\citenamefont{Adare et~al.}(2010{\natexlab{b}})}]{Adare:2009js}
\bibinfo{author}{\bibfnamefont{A.}~\bibnamefont{Adare}} \bibnamefont{et~al.}
  (\bibinfo{collaboration}{PHENIX Collaboration}), \bibinfo{journal}{Phys.
  Rev.} \textbf{\bibinfo{volume}{D82}}, \bibinfo{pages}{012001}
  (\bibinfo{year}{2010}{\natexlab{b}}).

\bibitem[{\citenamefont{Eskola et~al.}(2009)\citenamefont{Eskola, Paukkunen,
  and Salgado}}]{Eskola:2009uj}
\bibinfo{author}{\bibfnamefont{K.~J.} \bibnamefont{Eskola}},
  \bibinfo{author}{\bibfnamefont{H.}~\bibnamefont{Paukkunen}},
  \bibnamefont{and} \bibinfo{author}{\bibfnamefont{C.~A.}
  \bibnamefont{Salgado}}, \bibinfo{journal}{JHEP}
  \textbf{\bibinfo{volume}{04}}, \bibinfo{pages}{065} (\bibinfo{year}{2009}).

\bibitem[{\citenamefont{Vogt}(2005)}]{Vogt:2004dh}
\bibinfo{author}{\bibfnamefont{R.}~\bibnamefont{Vogt}}, \bibinfo{journal}{Phys.
  Rev.} \textbf{\bibinfo{volume}{C71}}, \bibinfo{pages}{054902}
  (\bibinfo{year}{2005}).

\bibitem[{\citenamefont{Kharzeev and Tuchin}(2004)}]{Kharzeev:2003sk}
\bibinfo{author}{\bibfnamefont{D.}~\bibnamefont{Kharzeev}} \bibnamefont{and}
  \bibinfo{author}{\bibfnamefont{K.}~\bibnamefont{Tuchin}},
  \bibinfo{journal}{Nucl. Phys.} \textbf{\bibinfo{volume}{A735}},
  \bibinfo{pages}{248} (\bibinfo{year}{2004}).

\bibitem[{\citenamefont{Klein and Vogt}(2003)}]{Klein:2003dj}
\bibinfo{author}{\bibfnamefont{S.~R.} \bibnamefont{Klein}} \bibnamefont{and}
  \bibinfo{author}{\bibfnamefont{R.}~\bibnamefont{Vogt}},
  \bibinfo{journal}{Phys. Rev. Lett.} \textbf{\bibinfo{volume}{91}},
  \bibinfo{pages}{142301} (\bibinfo{year}{2003}).

\bibitem[{\citenamefont{Frankfurt et~al.}(2005)\citenamefont{Frankfurt, Guzey,
  and Strikman}}]{Frankfurt:2003zd}
\bibinfo{author}{\bibfnamefont{L.}~\bibnamefont{Frankfurt}},
  \bibinfo{author}{\bibfnamefont{V.}~\bibnamefont{Guzey}}, \bibnamefont{and}
  \bibinfo{author}{\bibfnamefont{M.}~\bibnamefont{Strikman}},
  \bibinfo{journal}{Phys. Rev.} \textbf{\bibinfo{volume}{D71}},
  \bibinfo{pages}{054001} (\bibinfo{year}{2005}).

\bibitem[{\citenamefont{Qiu et~al.}(2002)\citenamefont{Qiu, Vary, and
  Zhang}}]{Qiu:1998rz}
\bibinfo{author}{\bibfnamefont{J.-w.} \bibnamefont{Qiu}},
  \bibinfo{author}{\bibfnamefont{J.~P.} \bibnamefont{Vary}}, \bibnamefont{and}
  \bibinfo{author}{\bibfnamefont{X.-f.} \bibnamefont{Zhang}},
  \bibinfo{journal}{Phys. Rev. Lett.} \textbf{\bibinfo{volume}{88}},
  \bibinfo{pages}{232301} (\bibinfo{year}{2002}).

\bibitem[{\citenamefont{Kopeliovich et~al.}(2010)\citenamefont{Kopeliovich,
  Potashnikova, Pirner, and Schmidt}}]{Kopeliovich:2010nw}
\bibinfo{author}{\bibfnamefont{B.~Z.} \bibnamefont{Kopeliovich}},
  \bibinfo{author}{\bibfnamefont{I.~K.} \bibnamefont{Potashnikova}},
  \bibinfo{author}{\bibfnamefont{H.~J.} \bibnamefont{Pirner}},
  \bibnamefont{and} \bibinfo{author}{\bibfnamefont{I.}~\bibnamefont{Schmidt}}
  (\bibinfo{year}{2010}).

\bibitem[{\citenamefont{Johnson et~al.}(2002)}]{Johnson:2001xfa}
\bibinfo{author}{\bibfnamefont{M.~B.} \bibnamefont{Johnson}}
  \bibnamefont{et~al.}, \bibinfo{journal}{Phys. Rev.}
  \textbf{\bibinfo{volume}{C65}}, \bibinfo{pages}{025203}
  (\bibinfo{year}{2002}).

\bibitem[{\citenamefont{Linden~Levy}(2009)}]{LindenLevy:2009td}
\bibinfo{author}{\bibfnamefont{L.~A.} \bibnamefont{Linden~Levy}},
  \bibinfo{journal}{Nucl. Phys.} \textbf{\bibinfo{volume}{A830}},
  \bibinfo{pages}{353c} (\bibinfo{year}{2009}).

\end{thebibliography}

\end{document}